\documentclass{amsart}
\usepackage{enumitem}
\usepackage[all]{xy}
\usepackage[margin=1.00in, marginparwidth=1.5cm, marginparsep=0.5cm]{geometry}

\usepackage{amsmath, amssymb}
\usepackage{algorithm}
\usepackage{algorithmicx}
\usepackage{algpseudocode}
\usepackage{float}
\usepackage{amsmath, amssymb, algorithm, algpseudocode, fancyvrb, tcolorbox}

\usepackage{mathtools}
\usepackage{bbold}

\usepackage{verbatim}
\usepackage{amssymb}
\usepackage{array}
\usepackage{siunitx}
\usepackage{mathrsfs}
\usepackage{physics}
\usepackage{amsthm}
\usepackage{amsmath,amsfonts,graphicx,mathdots,cite,wrapfig}
\usepackage[colorlinks,linkcolor=blue,citecolor=blue,pagebackref,hypertexnames=false, breaklinks]{hyperref}
\usepackage{cite}
\usepackage{mwe}
\usepackage{caption}
\usepackage{marginnote}
\usepackage{tcolorbox}

\newtheorem{theorem}{Theorem}[section]

\newtheorem{proposition}[theorem]{Proposition}

\theoremstyle{definition}
\newtheorem{definition}[theorem]{Definition}
\newtheorem{example}[theorem]{Example}
\theoremstyle{remark}
\newtheorem{remark}[theorem]{Remark}

\numberwithin{equation}{section}

\DeclareMathOperator{\nn}{\mathbb{N}}

\begin{document}



\title{Hierarchical Minimum Variance Portfolios: A Theoretical and Algorithmic Approach}




\author{Gamal Mograby}
\address{Department of Mathematical Sciences, University of Cincinnati, USA}
\email{mograbgl@ucmail.uc.edu}

\subjclass[2010]{q-fin.PM,q-fin.CP,q-fin.MF}

\date{\today}


\begin{abstract}
We introduce a novel approach to portfolio optimization that leverages hierarchical graph structures and the Schur complement method to systematically reduce computational complexity while preserving full covariance information. Inspired by López de Prado’s hierarchical risk parity and Cotton’s Schur complement methods, our framework models the covariance matrix as an adjacency-like structure of a hierarchical graph. We demonstrate that portfolio optimization can be recursively reduced across hierarchical levels, allowing optimal weights to be computed efficiently by inverting only small submatrices regardless of portfolio size. Moreover, we translate our results into a recursive algorithm that constructs optimal portfolio allocations. Our results reveal a transparent and mathematically rigorous connection between classical Markowitz mean-variance optimization, hierarchical clustering, and the Schur complement method.

\end{abstract}

\maketitle

\tableofcontents

\section{Introduction}

The optimization of financial portfolio allocation remains a cornerstone of investment strategy, continuously evolving to address the complexities of modern financial markets. Since Markowitz's seminal work on Modern Portfolio Theory \cite{Markowitz1952}, researchers have sought to develop methods that balance return maximization with risk minimization. However, classical portfolio optimization models, including mean-variance optimization, suffer from several practical limitations, such as sensitivity to estimation errors, instability in covariance matrix inversion, and poor out-of-sample performance \cite{michaud1998, lopezdeprado2016}.  

To address these challenges, recent advances in computational finance have introduced alternative frameworks that enhance portfolio stability and diversification. One such approach is Hierarchical Risk Parity, which leverages clustering techniques to construct more robust portfolios by mitigating estimation noise while preserving diversification benefits \cite{lopezdeprado2016}. Similarly, risk-budgeting methods, such as those utilizing Non-Negative Matrix Factorization, further improve stability by decomposing asset correlations into interpretable factors, ensuring a well-distributed risk exposure across portfolio components \cite{spilak2023nmf}. Beyond these techniques, reinforcement learning and stochastic modeling have emerged as promising approaches to dynamic asset allocation. The Multi-Armed Bandit framework enables portfolios to adapt to changing market conditions, while the \text{Multifractal Model of Asset Returns} captures non-Gaussian statistical patterns in financial time series, enhancing predictive capabilities beyond traditional econometric models \cite{mandelbrot1997variation, calvet2001forecasting}.

In this work, we focus on hierarchical portfolios pioneered by López de Prado, particularly the use of \text{graph-based techniques} to reorganize covariance matrices. Several researchers, including \cite{raffinot2017}, have extended this approach through Hierarchical Clustering-Based Asset Allocation, while the Hierarchical Equal Risk Contribution model further refines risk allocation by incorporating advanced risk measures \cite{raffinot2018}. While graph-based approaches to portfolio optimization are not new, they remain highly relevant. Mantegna’s correlation networks \cite{mantegna1999} introduced early applications of graph theory in financial markets, and more recent frameworks, such as those developed by \cite{cajas2023graph}, leverage the Minimum Spanning Tree and Triangulated Maximally Filtered Graph to enhance asset selection and diversification \cite{cajas2023graph}. More recently, Cotton applied Schur Complement methods to portfolio allocation, modifying covariance matrices to account for hierarchical structures \cite{cotton2024}. These methods reveal deeper mathematical connections between optimization and hierarchy.  

\begin{figure}[!htb]
        \centering
        \includegraphics[scale=0.6]{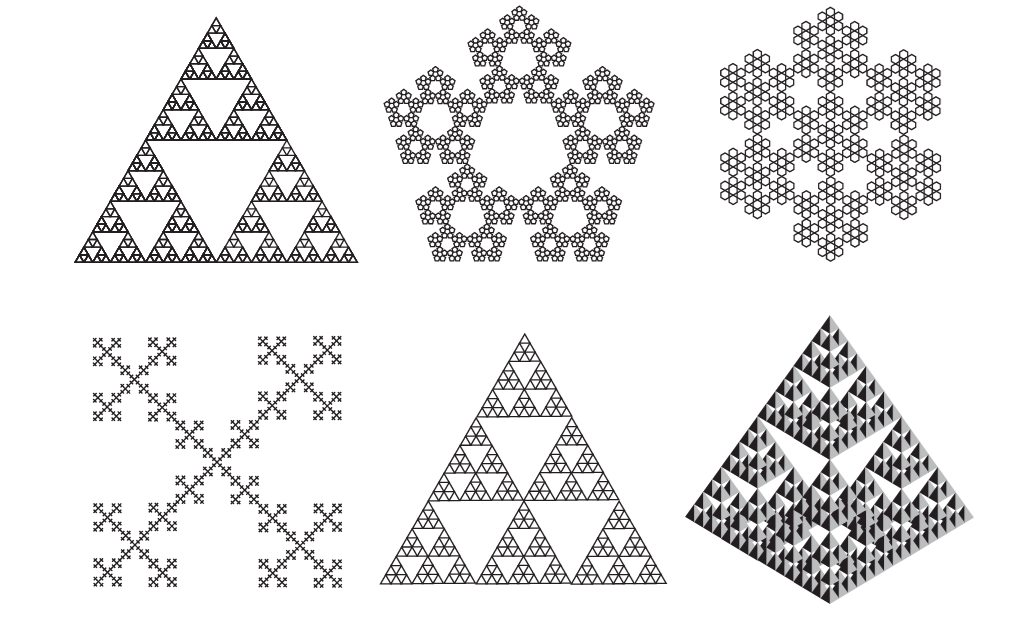}
    \caption{Examples of fractal-type graphs. These graphs are relevant to our work as they serve as templates and toy models for constructing hierarchical portfolios, demonstrating how self-similar structures can be effectively utilized for portfolio organization and risk management. The figure is adapted from Hino \cite{hino2008martingale}.}
	\label{fig:fractals}
\end{figure}

We propose an alternative approach to hierarchical portfolio construction, inspired by—but distinct from—the work of López de Prado \cite{lopezdeprado2016}, while also providing a natural setting in which Cotton’s ideas \cite{cotton2024} emerge organically. Our methodology relates the covariance matrix of a portfolio to a weighted graph, and if the graph belongs to the class of hierarchical graphs, we define the portfolio as a \textit{hierarchical portfolio}.

\textit{What Do We Mean by Hierarchical Graphs?} Hierarchical graphs are a recursively constructed sequence of graphs \( G_1, G_2, G_3, \dots \), where each level-\(\ell\) graph \( G_{\ell} \) is obtained by systematically combining multiple copies of the lower-level graph \( G_{\ell-1} \). This recursive structure is frequently used in applications where self-similarity and nested properties provide computational advantages. In mathematics, such sequences of graphs are often employed to approximate fractals.
For example, the top-left fractal-type graph in Figure~\ref{fig:fractals} approximates a fractal known as the Sierpiński gasket. In Figure~\ref{fig:sierpinski_graphs}, we illustrate the first four graphs in the hierarchical sequence that approximate the Sierpiński gasket. These graphs are constructed by iteratively combining smaller building blocks. The process in Figure~\ref{fig:sierpinski_graphs} starts with the simplest graph, denoted as \( G_1 \), and progresses by assembling multiple copies at each higher level. As the level \(\ell\) increases, the graph expands, adding more nodes and edges, thereby increasingly resembling a fractal-like structure.

In his work, López de Prado modeled the covariance matrix as a complete (fully connected) graph for a portfolio of 50 assets, as shown in Exhibit 2 of his paper \cite{lopezdeprado2016}. In a complete graph, every asset (represented as a node) is connected to every other asset, meaning that all assets are considered potential substitutes for one another. However, this lack of hierarchy in the complete graph implies that each node is treated as equivalent to any other, failing to capture the structural dependencies within financial markets. In reality, financial assets exhibit natural groupings based on industry sectors, geographic regions, or market capitalizations.

To introduce hierarchy, López de Prado proposed replacing the complete graph with tree-based graphs, which impose an organized clustering of assets. By incorporating hierarchical structures, tree-based graphs reduce complexity while preserving meaningful relationships among assets.

A key tradeoff in using tree-based graphs is that trees, by definition, do not contain cycles, meaning that certain correlation relationships between assets (nodes) are lost in this simplification. Moreover, constructing the tree itself requires defining a distance metric, which transforms correlation data into a form suitable for distance interpretation. This transformation inherently results in information loss, as it forces correlation structures to be interpreted through a metric that satisfies the mathematical properties of a distance function.

\textbf{Our Approach:} We will adhere to López de Prado’s initial idea of modeling a portfolio as a graph, similar to the complete graph representation in Exhibit 2 of his paper \cite{lopezdeprado2016}. However, instead of constructing trees, we propose a different step: our portfolios will be built from the beginning such that the covariance structure of their assets forms a hierarchical graph rather than a complete graph.

To understand this, we model portfolios using weighted graphs, where nodes represent assets, and edges are weighted by the covariance between these assets. Specifically, two nodes are connected by an edge if their corresponding assets are correlated, and the edge is assigned a weight equal to their covariance. Conversely, two nodes remain unconnected if the corresponding assets are uncorrelated. 

Simply put, we treat the covariance matrix as an adjacency-like matrix of a weighted graph. If all assets in the portfolio are pairwise correlated, then the covariance matrix corresponds to the adjacency matrix of a complete graph. However, in our approach, we construct the portfolio by selecting assets such that their covariance matrix describes the adjacency matrix of a hierarchical graph. 

For simplicity, in this paper, we focus on a specific class of hierarchical graphs—Sierpiński graphs. In Section~\ref{sec:GraphRepPortfolio}, we provide a detailed explanation of the relationship between Sierpiński graphs and portfolio covariance structures, formalizing how these graphs can serve as templates for hierarchical portfolio construction. While our discussion primarily centers on Sierpiński graphs, the ideas we introduce extend naturally to other families of hierarchical graphs and are not limited to this particular structure.

\begin{figure}[!htb]
        \centering
        \includegraphics[scale=0.9]{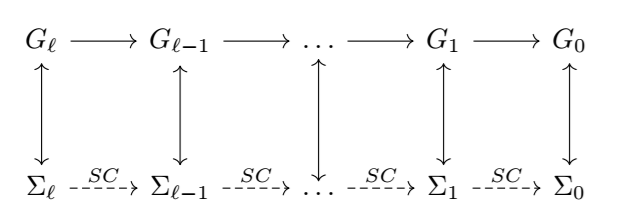}
        \caption{This figure illustrates the stepwise reduction of a hierarchical portfolio using the Schur complement ($SC$). At each level, the covariance matrix \( \Sigma_{\ell} \) of the portfolio, associated with a hierarchical graph \( G_{\ell} \), is transformed into an effective covariance matrix \( \Sigma_{\ell-1} \) corresponding to the reduced graph \( G_{\ell-1} \). This iterative process continues until reaching the base level, where the portfolio structure is represented by a simple graph \( G_0 \).}

	\label{fig:hierachyWithSigmas}
\end{figure}

The interpretation of the covariance matrix as an adjacency-like matrix of a graph can be applied generally, so why do we restrict our approach to hierarchical graphs? This is the focus of Section~\ref{sec:HierachyGraph}. Hierarchical graphs naturally reflect the grouping and subgrouping of assets in real-world portfolios, capturing structured dependencies within financial markets. Additionally, they provide significant mathematical advantages, which we summarize through our main contributions: Theorem~\ref{thm:mainResult} and Theorem~\ref{thm:mostimportant}.  

To summarize our contribution, let \( G_0, G_1, \dots, G_\ell \) represent the first \( \ell +1\) graphs in a sequence of hierarchical graphs. Suppose we construct a portfolio where the assets are selected such that their covariance matrix \( \Sigma_\ell \) corresponds to an adjacency-like matrix of \( G_\ell \). Our result, Theorem~\ref{thm:mostimportant}, states that solving the optimization problem for portfolio weights, i.e., finding  
\[
w_\ell = \Sigma_\ell^{-1} 1,
\]
can be recursively reduced to an equivalent optimization problem on the lower-level graph \( G_{\ell-1} \), expressed as  
\[
w_{\ell-1} = \Sigma_{\ell-1}^{-1} \gamma_{\ell-1}.
\]
Here, \( \Sigma_{\ell-1} \) represents the adjacency-like covariance matrix of the reduced graph \( G_{\ell-1} \) in the hierarchy. This process can be iteratively applied, reducing the optimization problem step by step until we reach the base graph \( G_0 \), where the final weights can be computed efficiently.  

We emphasize that this approach reduces computational complexity while preserving the complete covariance information. The key idea behind our framework is that \( \Sigma_{k-1} \) is obtained by applying the Schur complement to \( \Sigma_k \). Consequently, in a sequence of hierarchical graphs, the Schur complement serves as a level-reduction operator, effectively lowering the hierarchy by one level at each step, see Figure \ref{fig:hierachyWithSigmas}.  

For concreteness, we demonstrate our approach using the family of Sierpiński graphs. Regardless of how large the portfolio is, as long as its covariance matrix corresponds to an adjacency-like matrix of a Sierpiński graph, there is no need to compute the full inverse of \( \Sigma_\ell \). Instead, we show that the optimal portfolio weights can be computed efficiently by inverting only a series of \( 3 \times 3 \) matrices, a process that remains computationally feasible independent of portfolio size.

In Theorem~\ref{thm:mainResult}, we establish Cotton's idea \cite{cotton2024} of unifying the three approaches: Markowitz's mean-variance optimization, López de Prado's hierarchical risk parity, and the Schur complement method. Specifically, we show that the minimization of portfolio variance at two successive levels in the hierarchy can be systematically connected via the Schur complement method.

In Section~\ref{sec:Algo}, we translate our results into a recursive algorithm for computing optimal portfolio weights. In this algorithm, depending on the chosen hierarchical graph used as a template for portfolio construction, the computational complexity is significantly reduced. Specifically, the problem is reduced to inverting several low-dimensional matrices, where the matrix size corresponds to the number of nodes in \( G_0 \) within the hierarchy. For example, in our Sierpiński hierarchy, if we start with \( \Sigma_\ell \), it is an \( n \times n \) matrix, where  

\[
n = \frac{1}{2} (3^{\ell+1} + 3).
\]
\\
Thus, for large \( \ell \), we are dealing with large portfolios. However, in our algorithm, the computation only requires inverting \( 3 \times 3 \) matrices, as the number \( 3 \) corresponds to the number of nodes in \( G_0 \), which is a triangle graph in the Sierpiński graph hierarchy.

\section{Graph Representation of Portfolios}
\label{sec:GraphRepPortfolio}

\begin{figure}[!htb]
        \centering
        \includegraphics[scale=0.55]{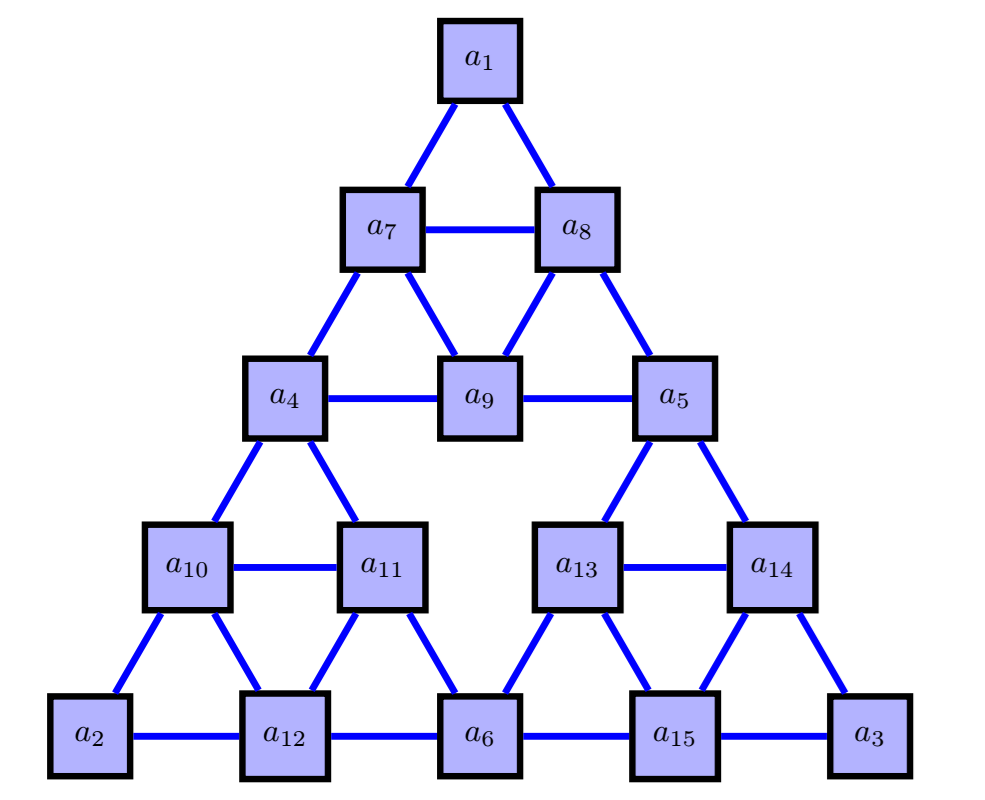}
    	\caption{Graph representation of a portfolio consisting of 15 assets. Each node represents an asset, and edges between nodes are weighted by the covariance between asset returns. Assets with nonzero covariance are connected by edges, while uncorrelated assets remain disconnected.}
	\label{fig:portfolioLevel2}
\end{figure}

In this section, we introduce a graph-theoretical approach to portfolio analysis, offering a structured and visual representation of a portfolio. By leveraging graph structures, this method uncovers hidden patterns and provides an intuitive framework for understanding diversification and clustering within a portfolio. We consider a market consisting of \( n \) assets, denoted as \( a_1, \dots, a_n \), with returns at a given time represented by \( R_1, \dots, R_n \). An investor allocating a fraction \( w_i \) of their total wealth to asset \( a_i \) constructs a portfolio with a total return given by:
\begin{equation}
\label{eq:returns}
R_{\text{tot}} = \sum_{i=1}^{n} w_i R_i.
\end{equation}
subject to the normalization condition:
\begin{equation}
\label{eq:weights}
\sum_{i=1}^{n} w_i = 1,
\end{equation}
which ensures that the portfolio allocations sum to one. Such a portfolio can be represented as a weighted graph, where each security corresponds to a node, and the edges between nodes are weighted according to the covariance between the respective returns (see for instance Figure~\ref{fig:portfolioLevel2}). The covariance structure of the portfolio is naturally encoded in the graph’s adjacency matrix: if two assets are uncorrelated, no edge exists between their corresponding nodes.

To develop a formal framework for graph-based portfolio analysis, we begin with the definition of a weighted graph.

\begin{definition}
\label{def:PortfolioGraph}
A \textit{weighted graph} is a triplet \( G = (V, E, \Sigma) \), where:
\begin{itemize}
    \item \( V = \{a_1, a_2, \dots, a_n\} \) is the set of nodes. In our context, each node represents an individual security in a portfolio.
    \item \( E \subseteq V \times V \) represents the set of edges, defining relationships between nodes. Two nodes are adjacent if they share an edge. In the portfolio model, an edge exists between two nodes if the returns of their corresponding assets exhibit a nonzero covariance.
    \item The function \( \Sigma: V \times V \to \mathbb{R} \) is a weight function that assigns a numerical value, called the weight, to both nodes and edges. In our context, edge weights are derived from the covariance matrix of asset returns. Specifically, for any two nodes \( a_i \) and \( a_j \), the edge \( (a_i, a_j) \) exists (i.e., they are adjacent) if their covariance is nonzero, with the weight given by:
\begin{equation*}
\Sigma(a_i, a_j) = \text{Cov}(R_i, R_j) = \sigma_{ij}.
\end{equation*}
If two assets are uncorrelated, meaning \( \sigma_{ij} = 0 \), no edge is present in the graph, i.e., \( (a_i, a_j) \notin E \). For self-loops, i.e., the case where \( i = j \), the function \( \Sigma \) assigns a weight to the node \( a_i \) corresponding to the variance of its returns:
\begin{equation*}
\Sigma(a_i, a_i) = \text{Cov}(R_i, R_i) = \text{Var}(R_i) = \sigma_{ii}= \sigma_{i}^2.
\end{equation*}
\end{itemize}
\end{definition}
\begin{example}
\label{ex:Level2Example}
We consider a portfolio consisting of 15 assets, denoted as \( a_1, a_2, \dots, a_{15} \), with its graph representation shown in Figure~\ref{fig:portfolioLevel2}. Each asset is represented as a node in a weighted graph, where edges are assigned weights based on the covariance matrix\footnote{A detailed explanation of the subscript notation in $\mathbf{\Sigma}_2$ and the block structure of the matrix can be found in next sections.} $\mathbf{\Sigma}_2$ in (\ref{eq:SigmaLevel2}). Each diagonal entry \( \sigma_{ii} \) represents the variance of asset \( a_i \). For example, the variances of the first three assets are \( \sigma_{11} = 7 \), \( \sigma_{22} = 13 \), and \( \sigma_{33} = 8 \), and so on. The off-diagonal elements \( \sigma_{ij} \) capture the covariance between assets \( a_i \) and \( a_j \). If \( \sigma_{ij} \neq 0 \), an edge is formed between the corresponding nodes, with the covariance value determining the edge weight. For example, \( \sigma_{17} = -3 \), \( \sigma_{47} = 4 \), and \( \sigma_{58} = 1 \), and so on. This translates into the following edge relationships in Figure~\ref{fig:portfolioLevel2}:
\begin{itemize}
    \item Asset \( a_1 \) is connected to asset \( a_7 \) with a weight of \(-3\).
    \item Asset \( a_4 \) is connected to asset \( a_7 \) with a weight of \( 4 \).
    \item Asset \( a_5 \) is connected to asset \( a_8 \) with a weight of \( 1 \).
\end{itemize}
Conversely, if \( \sigma_{ij} = 0 \), the assets are uncorrelated, meaning no edge is formed between them. For instance, \( \sigma_{12} = 0 \), \( \sigma_{13} = 0 \), and \( \sigma_{46} = 0 \), compare Figure~\ref{fig:portfolioLevel2}. This implies the following:
\begin{itemize}
    \item Asset \( a_1 \) and asset \( a_2 \) are not adjacent nodes.
    \item Asset \( a_1 \) and asset \( a_3 \) are not adjacent nodes.
    \item Asset \( a_4 \) and asset \( a_6 \) are and not adjacent nodes.
\end{itemize}
\end{example}
\begin{equation}
\label{eq:SigmaLevel2}
\mathbf{\Sigma}_{2} =
\left(
\begin{array}{cccccc|ccccccccc}
7 & 0 & 0 & 0 & 0 & 0 & -3 & 0 & 0 & 0 & 0 & 0 & 0 & 0 & 0\\
0 & 13 & 0 & 0 & 0 & 0 & 0 & 0 & 0 & -3 & 0 & 3 & 0 & 0 & 0\\
0 & 0 & 8 & 0 & 0 & 0 & 0 & 0 & 0 & 0 & 0 & 0 & 0 & -5 & 0\\
0 & 0 & 0 & 7 & 0 & 0 & 4 & 0 & 0 & -3 & -2 & 0 & 0 & 0 & 0\\
0 & 0 & 0 & 0 & 12 & 0 & 0 & 0 & 1 & 0 & 0 & 0 & -2 & 0 & 0\\
0 & 0 & 0 & 0 & 0 & 7 & 0 & 0 & 0 & 0 & 0 & -1 & 0 & 0 & 1\\
\hline
-3 & 0 & 0 & 4 & 0 & 0 & \multicolumn{1}{|c}{9} & -1 & \multicolumn{1}{c|}{4} & 0 & 0 & 0 & 0 & 0 & 0\\
0 & 0 & 0 & 0 & 0 & 0 & \multicolumn{1}{|c}{-1} & 11 & \multicolumn{1}{c|}{-6} & 0 & 0 & 0 & 0 & 0 & 0\\
0 & 0 & 0 & 0 & 1 & 0 & \multicolumn{1}{|c}{4} & -6 & \multicolumn{1}{c|}{12} & 0 & 0 & 0 & 0 & 0 & 0\\
\cline{7-12}
0 & -3 & 0 & -3 & 0 & 0 & 0 & 0 & 0 & \multicolumn{1}{|c}{13} & -1 & \multicolumn{1}{c|}{-3} & 0 & 0 & 0\\
0 & 0 & 0 & -2 & 0 & 0 & 0 & 0 & 0 & \multicolumn{1}{|c}{-1} & 8 & \multicolumn{1}{c|}{-1} & 0 & 0 & 0\\
0 & 3 & 0 & 0 & 0 & -1 & 0 & 0 & 0 & \multicolumn{1}{|c}{-3} & -1 & \multicolumn{1}{c|}{9} & 0 & 0 & 0\\
\cline{10-15}
0 & 0 & 0 & 0 & -2 & 0 & 0 & 0 & 0 & 0 & 0 & 0 & \multicolumn{1}{|c}{9} & 1 & \multicolumn{1}{c|}{3}\\
0 & 0 & -5 & 0 & 0 & 0 & 0 & 0 & 0 & 0 & 0 & 0 & \multicolumn{1}{|c}{1} & 12 &  \multicolumn{1}{c|}{-1}\\
0 & 0 & 0 & 0 & 0 & 1 & 0 & 0 & 0 & 0 & 0 & 0 & \multicolumn{1}{|c}{3} & -1 &  \multicolumn{1}{c|}{8} \\
\cline{13-15}
\end{array}
\right)
\end{equation}
\text{ }

Having established how the graph structure naturally reflects the covariance relationships within a portfolio, another advantage of this representation is that portfolio weights (\ref{eq:weights}) and asset returns (\ref{eq:returns}) can also be naturally interpreted as functions defined on the set of nodes. Specifically, we define the \textit{portfolio weight function} \( w: V \to \mathbb{R} \) and the \textit{asset return function} \( R: V \to \mathbb{R} \), which assign a weight and a return, respectively, to each node in the graph. That is, for each asset \( a_i \) in the portfolio:
\begin{equation*}
w(a_i) = w_i, \quad R(a_i) = R_i.
\end{equation*}
Since each vertex \( a_i \) represents an asset in the portfolio, By establishing a fixed ordering of the nodes, we can represent these functions as vectors:
\begin{equation*}
\mathbf{w} = (w_1, w_2, \dots, w_n)^t, \quad \quad \mathbf{R} = (R_1, R_2, \dots, R_n)^t.
\end{equation*}
Using this notation, the total portfolio return (\ref{eq:returns}) is expressed as:
\begin{equation*}
R_{\text{tot}} = \mathbf{w}^t \mathbf{R} = \sum_{i=1}^{n} w_i R_i.
\end{equation*}
When selecting a portfolio, an investor has several approaches to optimizing their investment. In this work, we focus on minimizing risk. A key measure of portfolio risk is the \textit{portfolio variance}, which quantifies uncertainty based on asset dependencies. We denote the \text{covariance matrix} by:
\begin{equation*}
\boldsymbol{\Sigma} = (\sigma_{ij})_{i,j}, \quad \sigma_{ij} = \text{Cov}(R_i, R_j).
\end{equation*}
The total portfolio variance is then given by:
\begin{equation}
\label{eq:optPortVector}
\sigma_{\text{tot}}^2 = \mathbf{w}^t \boldsymbol{\Sigma} \mathbf{w} = \sum_{i=1}^{n} \sum_{j=1}^{n} w_i  \sigma_{ij} w_j.
\end{equation}
Since portfolio variance \( \sigma_{\text{tot}}^2 \) determines risk exposure, minimizing it is a central objective in portfolio optimization. We seek the optimal weight vector \( \mathbf{w} \) that minimizes portfolio variance while ensuring that the total portfolio weight sums to one, i.e.
\begin{equation*}
\min_{\mathbf{w}} \Big\{ \mathbf{w}^t \boldsymbol{\Sigma} \mathbf{w} \ \Big | \ \sum_{i=1}^{n} w_i = 1 \ \Big\}
\end{equation*}
Under the assumption that the covariance matrix is invertible, the minimum variance portfolio is computed using:
\begin{equation}
\label{eq:Optimal_w}
    \mathbf{w}^* = \frac{\boldsymbol{\Sigma}^{-1} \mathbf{1}}{\mathbf{1}^t \boldsymbol{\Sigma}^{-1} \mathbf{1}}.
\end{equation}
where \( \mathbf{1} \) is an \( n \)-dimensional column vector of ones, $\mathbf{1} = (1,1,\dots,1)^t \in \mathbb{R}^n$. 

Note that the covariance matrix \( \Sigma \) is positive semi-definite, meaning it is \text{invertible if and only if} it is positive definite. Under the assumption that \( \Sigma \) is positive definite, the portfolio return variance \( R_{\text{tot}} \) induces an inner product:

\begin{equation}
\label{eq:innerProductNot}
\langle \mathbf{w}, \mathbf{w} \rangle_{\Sigma} = \mathbf{w}^t \Sigma \mathbf{w}.
\end{equation}
\\
In this sense, minimizing \( R_{\text{tot}} \) is equivalent to minimizing the \textit{length (norm)} of \( \mathbf{w} \) induced by the \( \Sigma \), i.e. $\vert \vert \mathbf{w} \vert \vert_{\Sigma} = \sqrt{\langle \mathbf{w}, \mathbf{w} \rangle_{\Sigma}}$ .
\begin{example}[Continuation of Example \ref{ex:Level2Example}]
\label{ex:Level2Example1}
Using (\ref{eq:Optimal_w}), the optimal portfolio weights are computed via direct inversion of the covariance matrix (\ref{eq:SigmaLevel2}). We obtain,
\begin{equation}
\label{eq:optimalWeightsConcrete}
\small
w^* \approx
\left( 0.04, \  0.035, \  0.097, \  0.158, \  0.032, \  0.064, -0.041, \  0.079, \  0.084, \  0.105, \  0.114, \  0.088, \  0.03, \  0.075, \ 0.04\right)^t
\end{equation}
The corresponding portfolio variance, calculated using (\ref{eq:optPortVector}), is $\sigma_{\text{tot}}^2 \approx 0.403$.
\end{example}
Computing the optimal weights requires inverting the covariance matrix \( \mathbf{\Sigma} \). As we will see below, leveraging a hierarchical structure allows us to break down the portfolio into smaller, more manageable components. In particular, we will show that under a well-defined hierarchy, the optimal portfolio weights can be determined by inverting only matrices of size \( 3 \times 3 \), \textbf{regardless} of the size of the original portfolio, as long as it adheres to the hierarchical structure.

\section{Hierarchical Graphs for Portfolio Construction}
\label{sec:HierachyGraph}

In this section, we define \textit{hierarchical graphs} as a sequence of recursively constructed graphs \( G_1, G_2, G_3, \dots \), where each level-\( \ell \) graph \( G_{\ell} \) is formed by systematically combining multiple copies of the lower-level graph \( G_{\ell-1} \). While various hierarchical graphs can be used, we focus on \textit{Sierpiński graphs} for simplicity. However, the underlying principles of our results apply to other hierarchical structures as well.

\begin{figure}[h]
\centering
\includegraphics[scale=1]{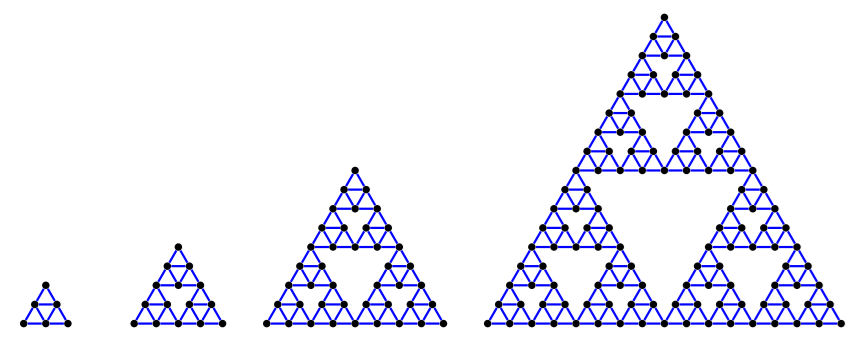}
\caption{From left to right, the sequence of graphs represents \textit{Level 1} to \textit{Level 4} of the Sierpiński graphs. Each level is constructed by combining three copies of the previous level. For the construction details, the reader is referred to \cite{Teplyaev1998, KigamiBook2001}.}
	\label{fig:sierpinski_graphs}
\end{figure}

Figure~\ref{fig:sierpinski_graphs} illustrates the construction of Sierpiński graphs. The process begins with the leftmost graph, known as the \textit{Level 1} Sierpiński graph, denoted \( G_1 \). By combining three copies of \( G_1 \), we obtain the \textit{Level 2} graph, \( G_2 \). This recursive pattern continues, where three copies of \( G_2 \) form \( G_3 \), and so on. Notably, \( G_1 \) serves as the fundamental building block for all these graphs. For example, \( G_2 \) consists of three copies of \( G_1 \), while \( G_3 \) contains nine copies of \( G_1 \), and so forth. In this sense, \( G_1 \) acts as a \textit{base cluster of assets} used to construct a portfolio.
Within each graph \( G_{\ell} \), we distinguish two types of nodes:
\begin{itemize}
\item \textbf{Junction nodes}: These are the \textit{corner nodes} of each copy of \( G_1 \) in \( G_{\ell} \). They serve as connection points between different copies of \( G_1 \), effectively acting as separators of the base clusters.
    \item \textbf{Interior nodes}: These are nodes that remain entirely within a single copy of \( G_1 \) and do not contribute to connecting different clusters.  
\end{itemize}

\begin{example}[Continuation of Example \ref{ex:Level2Example}]
The graph in Figure~\ref{fig:portfolioLevel2} represents a Level 2 Sierpiński graph. It is the second leftmost graph in Figure~\ref{fig:sierpinski_graphs}. The subscript in the covariance matrix \( \Sigma_2 \) indicates the corresponding hierarchical level. The junction nodes (assets) are \( \{a_1, a_2, a_3, a_4, a_5, a_6\} \), which are uncorrelated with each other. Their primary role is to enhance diversification by connecting different clusters while minimizing direct correlations. The interior nodes (assets) are grouped as \( \{\{a_7, a_8, a_9\}, \{a_{10}, a_{11}, a_{12}\}, \{a_{13}, a_{14}, a_{15}\}\} \), where each group of three assets is correlated, forming a cohesive cluster within the portfolio. To structure the covariance matrix efficiently, we first order the junction assets, followed by the interior assets, grouped cluster by cluster. This ordering results in a block-structured covariance matrix:
\begin{equation*} 
\mathbf{\Sigma}_2 =
\begin{pmatrix}
\mathbf{T}_2 & \mathbf{J}_2^t \\
\mathbf{J}_2 & \mathbf{X}_2
\end{pmatrix}
\end{equation*}
Comparing with \eqref{eq:SigmaLevel2}, we identify the following components:
\begin{itemize}
    \item \( \mathbf{T}_2 \) represents the covariance submatrix corresponding to the junction assets. It is diagonal, reflecting the assumption that the junction assets are uncorrelated with each other.
    \item \( \mathbf{X}_2 \) represents the covariance among interior assets. It consists of three block matrices of size \( 3 \times 3 \), corresponding to the covariance matrices of the interior assets in each cluster. Note that there are three sub-blocks in \( \mathbf{X} \) because the Level 2 Sierpiński graph consists of three clusters.
    \item \( \mathbf{J}_2\) encodes the interdependencies between the interior and junction assets.
\end{itemize}
\end{example}

Now, we consider \( G_\ell \), the Sierpiński graph at level \( \ell \), where we decompose the nodes into junction nodes and interior nodes, as explained above. We are interested in finding the optimal portfolio weights (\ref{eq:Optimal_w}). We decompose the weight vector \( \mathbf{w} \) into two components:  
\begin{equation}
\mathbf{w} =
\begin{pmatrix}
\mathbf{w}_{\text{jun}} \\
\mathbf{w}_{\text{in}}
\end{pmatrix},
\quad \mathbf{1}^t_{\text{jun}} \mathbf{w}_{\text{jun}} + \mathbf{1}^t_{\text{in}} \mathbf{w}_{\text{in}} = 1.
\end{equation}
where \( \mathbf{w}_{\text{jun}} \) represents the weights assigned to junction nodes, and \( \mathbf{w}_{\text{in}} \) represents the weights assigned to interior nodes. Similarly, \( \mathbf{1}_{\text{jun}} \) and \( \mathbf{1}_{\text{in}} \) are vectors of ones corresponding to the junction and interior nodes, respectively. The following result states that to compute the optimal weight vector \( \mathbf{w} \), all we need is the weights on the junction nodes, \( \mathbf{w}_{\text{jun}} \).

\begin{proposition}
\label{prop:Schur}
Let \( \mathbf{y} \) be a vector defined on all nodes, and decompose it as  
$\mathbf{y}^t = \big(\mathbf{y}_{\text{jun}}^t, \ \mathbf{y}_{\text{in}}^t \big)$. Define the following functions:
\begin{enumerate}
\item For a vector \( \mathbf{v} \) defined on the junction nodes, define  
$
   h(\mathbf{v}, \mathbf{y}) = \mathbf{X}^{-1} \mathbf{y}_{\text{in}} - \mathbf{X}^{-1} \mathbf{J} \mathbf{v}.
$
\item Define  
$
   g(\mathbf{y}) = \mathbf{y}_{\text{jun}} - \mathbf{J}^t \mathbf{X}^{-1} \mathbf{y}_{\text{in}}.
$
\end{enumerate}
Then, a vector \( \mathbf{w} \) satisfies the equation \( \mathbf{w} = \mathbf{\Sigma}^{-1} \mathbf{y} \) \text{if and only if} it can be written in the form
\begin{equation}
\label{eq:transJunToInt}
\mathbf{w} =
\begin{pmatrix}
\mathbf{w}_{\text{jun}} \\
h(\mathbf{w}_{\text{jun}}, \mathbf{y})
\end{pmatrix}.
\end{equation}
Furthermore, \( \mathbf{w}_{\text{jun}} \) satisfies the equation  
$\mathbf{w}_{\text{jun}} = \mathbf{S}(\Sigma)^{-1} g(\mathbf{y})$,
where \( \mathbf{S}(\Sigma) \) is the Schur complement of \( \mathbf{X} \) in \( \mathbf{\Sigma} \), given by $\mathbf{S}(\Sigma) = \mathbf{T} - \mathbf{J}^t \mathbf{X}^{-1} \mathbf{J}$.
\end{proposition}
\begin{proof}
We start with the assumption $\mathbf{w} = \mathbf{\Sigma}^{-1} \mathbf{y}$, which is equivalent to $\mathbf{\Sigma} \mathbf{w} =  \mathbf{y}$. Using the block structure of the covariance matrix \( \mathbf{\Sigma} \), we write:
\[
\begin{pmatrix}
\mathbf{T} & \mathbf{J}^t \\
\mathbf{J} & \mathbf{X}
\end{pmatrix}
\begin{pmatrix}
\mathbf{w}_{\text{jun}} \\
\mathbf{w}_{\text{in}}
\end{pmatrix}
=
\begin{pmatrix}
\mathbf{y}_{\text{jun}} \\
\mathbf{y}_{\text{in}}
\end{pmatrix}
\]
Expanding this system gives two equations. From the second equation $\mathbf{J} \mathbf{w}_{\text{jun}} + \mathbf{X} \mathbf{w}_{\text{in}} = \mathbf{y}_{\text{in}}$, we solve for \( \mathbf{w}_{\text{in}} \):
\[
\mathbf{w}_{\text{in}} = \mathbf{X}^{-1} \mathbf{y}_{\text{in}} - \mathbf{X}^{-1} \mathbf{J} \mathbf{w}_{\text{jun}} = h(\mathbf{w}_{\text{jun}},\mathbf{y}) 
\]
Now, substituting \( \mathbf{w}_{\text{in}} \) into the first equation 
$
\mathbf{T} \mathbf{w}_{\text{jun}} + \mathbf{J}^t (\mathbf{X}^{-1} \mathbf{y}_{\text{in}} - \mathbf{X}^{-1} \mathbf{J} \mathbf{w}_{\text{jun}}) = \mathbf{y}_{\text{jun}},
$
and rearranging, we obtain
\begin{equation*}
(\mathbf{T} - \mathbf{J}^t \mathbf{X}^{-1} \mathbf{J}) \mathbf{w}_{\text{jun}} = \mathbf{y}_{\text{jun}} - \mathbf{J}^t \mathbf{X}^{-1} \mathbf{y}_{\text{in}}.
\end{equation*}
By the definition of \( \mathbf{S}(\Sigma) \) and \( g(\mathbf{y}) \), we obtain
$
\mathbf{S}(\Sigma) \mathbf{w}_{\text{jun}} = g(\mathbf{y}).
$
Solving for \( \mathbf{w}_{\text{jun}} \) gives  
$
\mathbf{w}_{\text{jun}} = \mathbf{S}(\Sigma)^{-1} g(\mathbf{y}).
$
\end{proof}
We will determine the optimal weights (\ref{eq:Optimal_w}) by considering the case where \( \mathbf{w} = \mathbf{\Sigma}^{-1} \mathbf{1} \), corresponding to \( \mathbf{y} = \mathbf{1} \) in Proposition \ref{prop:Schur}. This proposition shows that once the weights on the junction nodes are determined, the weights for the interior nodes can be computed directly using the function \(  \mathbf{w}_{\text{in}} = h(\mathbf{w}_{\text{jun}},\mathbf{1}) \). This marks the first step in a \textit{divide-and-conquer} approach, where the optimization problem is systematically reduced to a smaller set of assets, specifically the junction nodes. To fully establish this framework, we now analyze how the inner product \( \langle \mathbf{w}, \mathbf{w} \rangle_{\Sigma} = \mathbf{w}^t \Sigma \mathbf{w} \) decomposes under this hierarchical structure.

\begin{proposition}
\label{prop:DecompOfVariance}
Let the function \( h(\mathbf{v}, \mathbf{y}) \) and the Schur complement \( \mathbf{S}(\Sigma) \) of \( \mathbf{X} \) in \( \mathbf{\Sigma} \) be defined as in Proposition \ref{prop:Schur}. Let \( \mathbf{w} \) and \( \mathbf{y} \) be vectors defined on all nodes. Then, the portfolio variance can be decomposed as   
\begin{align*}
\langle \mathbf{w}, \mathbf{w} \rangle_{\Sigma} =
& \ \langle \mathbf{w}_{\text{jun}}, \mathbf{w}_{\text{jun}} \rangle_{\mathbf{S}(\Sigma)} 
+ \mathbf{y}_{\text{in}}^t \mathbf{X}^{-1} \mathbf{y}_{\text{in}} \\
& + \langle \mathbf{w}_{\text{in}} - h(\mathbf{w}_{\text{jun}},\mathbf{y}), 
\mathbf{w}_{\text{in}} - h(\mathbf{w}_{\text{jun}},\mathbf{y})  \rangle_{\mathbf{X}}  + 2 \mathbf{y}_{\text{in}}^t \big( \mathbf{w}_{\text{in}} - h(\mathbf{w}_{\text{jun}},\mathbf{y}) \big).
\end{align*}
where we used the inner product notation for the variance introduced in (\ref{eq:innerProductNot}).
\end{proposition}

\begin{proof}
We begin by computing the inner product on the left-hand side:
\begin{equation}
\label{eq:comInnerProd}
\langle \mathbf{w}, \mathbf{w} \rangle_{\mathbf{\Sigma}}  = 
\mathbf{w}_{\text{jun}}^t \mathbf{T} \mathbf{w}_{\text{jun}} + 2 \mathbf{w}_{\text{jun}}^t \mathbf{J}^t \mathbf{w}_{\text{in}} + \mathbf{w}_{\text{in}}^t \mathbf{X} \mathbf{w}_{\text{in}}.
\end{equation}
Next, we compute the third term on the right-hand side:
\begin{equation*}
\mathbf{w}_{\text{in}}^t \mathbf{X} \mathbf{w}_{\text{in}}
+ 2 \mathbf{w}_{\text{jun}}^t \mathbf{J}^t \mathbf{w}_{\text{in}}
- 2 \mathbf{w}_{\text{in}}^t \mathbf{y}_{\text{in}}
+ \mathbf{y}_{\text{in}}^t \mathbf{X}^{-1} \mathbf{y}_{\text{in}}
- 2 \mathbf{w}_{\text{jun}}^t \mathbf{J}^t \mathbf{X}^{-1} \mathbf{y}_{\text{in}}
+ \mathbf{w}_{\text{jun}}^t \mathbf{J}^t \mathbf{X}^{-1} \mathbf{J} \mathbf{w}_{\text{jun}}.
\end{equation*}
We now compute the first term on the right-hand side:
\begin{equation*}
\mathbf{w}_{\text{jun}}^t \mathbf{T} \mathbf{w}_{\text{jun}} - \mathbf{w}_{\text{jun}}^t \mathbf{J}^t \mathbf{X}^{-1} \mathbf{J} \mathbf{w}_{\text{jun}}.
\end{equation*}
Combining these two expressions and simplifying their sum using (\ref{eq:comInnerProd}) gives the following:
\begin{equation*}
\langle \mathbf{w}, \mathbf{w} \rangle_{\mathbf{\Sigma}} 
+ \mathbf{y}_{\text{in}}^t \mathbf{X}^{-1} \mathbf{y}_{\text{in}}
- 2 \mathbf{y}_{\text{in}}^t \mathbf{w}_{\text{in}}
- 2 \mathbf{y}_{\text{in}}^t \mathbf{X}^{-1} \mathbf{J} \mathbf{w}_{\text{jun}}.
\end{equation*}
Rearranging the terms, we obtain:
\begin{equation*}
\langle \mathbf{w}, \mathbf{w} \rangle_{\mathbf{\Sigma}} 
- \mathbf{y}_{\text{in}}^t \mathbf{X}^{-1} \mathbf{y}_{\text{in}}
- 2 \mathbf{y}_{\text{in}}^t \mathbf{w}_{\text{in}}
- 2 \mathbf{y}_{\text{in}}^t \big( \mathbf{X}^{-1} \mathbf{J} \mathbf{w}_{\text{jun}} - \mathbf{X}^{-1} \mathbf{y}_{\text{in}} \big).
\end{equation*}
Finally, since $h(\mathbf{w}_{\text{jun}},\mathbf{y}) = \mathbf{X}^{-1} \mathbf{y}_{\text{in}} - \mathbf{X}^{-1} \mathbf{J} \mathbf{w}_{\text{jun}}$, the first and third terms on the right-hand side simplify to:
\begin{equation}
\langle \mathbf{w}, \mathbf{w} \rangle_{\mathbf{\Sigma}} 
- \mathbf{y}_{\text{in}}^t \mathbf{X}^{-1} \mathbf{y}_{\text{in}}
- 2 \mathbf{y}_{\text{in}}^t \big( \mathbf{w}_{\text{in}}
- h(\mathbf{w}_{\text{jun}},\mathbf{y}) \big).
\end{equation}
\end{proof}
The decomposition of \( \mathbf{w}^t \mathbf{\Sigma} \mathbf{w} \) in Proposition \ref{prop:DecompOfVariance} provides insight into how risk is distributed across junction and interior assets. When combined with Proposition \ref{prop:Schur}, this result justifies why solving for the junction node weights first is sufficient for optimization. Instead of working with the full covariance matrix, we can transform it into a smaller, reduced covariance structure using the Schur complement, which makes the problem more computationally efficient while preserving the key relationships between assets. The following theorem formalizes this observation.

\begin{theorem}
\label{thm:mainResult}
Let \( \mathbf{w} \) be defined as \( \mathbf{w} = \mathbf{\Sigma}^{-1} \mathbf{1} \). We decompose \( \mathbf{w} \) into junction and interior node weights
$
\mathbf{w} =
\begin{pmatrix}
\mathbf{w}_{\text{jun}},
\mathbf{w}_{\text{in}}
\end{pmatrix}^t
$
and compute the following:
\begin{enumerate}
    \item Compute the junction node weights by solving the following equation:
    \begin{align*}
    \mathbf{w}_{\text{jun}} = \mathbf{S}(\Sigma)^{-1}g(\mathbf{1}), \quad \quad g(\mathbf{1})= \mathbf{1}_{\text{jun}} - \mathbf{J}^t \mathbf{X}^{-1} \mathbf{1}_{\text{in}},
    \end{align*}
    where \( \mathbf{S}(\Sigma) = \mathbf{T} - \mathbf{J}^t \mathbf{X}^{-1} \mathbf{J} \) is the Schur complement.

    \item Once \( \mathbf{w}_{\text{jun}} \) is obtained, the interior node weights follow from:
    \[
    \mathbf{w}_{\text{in}} = h(\mathbf{w}_{\text{jun}}, \mathbf{1}), \quad \quad h(\mathbf{w}_{\text{jun}}, \mathbf{1}) = \mathbf{X}^{-1} \mathbf{1}_{\text{in}} - \mathbf{X}^{-1} \mathbf{J} \mathbf{w}_{\text{jun}}
    \]

    \item The vector \( \mathbf{w} \) minimizes \( \langle \mathbf{w}, \mathbf{w} \rangle_{\Sigma} \) if and only if the junction node weights \( \mathbf{w}_{\text{jun}} \) minimize \( \langle \mathbf{w}_{\text{jun}}, \mathbf{w}_{\text{jun}} \rangle_{\mathbf{S}(\Sigma)} \). Specifically, they differ by a positive constant term independent of \( \mathbf{w} \), given by:
    
\begin{equation}
\label{eq:variaceTwoLevels}
\langle \mathbf{w}, \mathbf{w} \rangle_{\Sigma} =
\langle \mathbf{w}_{\text{jun}}, \mathbf{w}_{\text{jun}} \rangle_{\mathbf{S}(\Sigma)}
+ \mathbf{1}_{\text{in}}^t \mathbf{X}^{-1} \mathbf{1}_{\text{in}}.
\end{equation}
\item Finally, we determine the minimum variance portfolio \( \mathbf{w}^* \), as defined in Equation (\ref{eq:Optimal_w}):
\[
\mathbf{w}^* =
\begin{pmatrix}
\mathbf{w}^{*}_{\text{jun}} \\[5pt]
\mathbf{w}^{*}_{\text{in}}
\end{pmatrix}
=
\frac{1}{\mathbf{1}_{\text{jun}}^t \mathbf{w}_{\text{jun}} + \mathbf{1}_{\text{in}}^t \mathbf{w}_{\text{in}}}
\begin{pmatrix}
\mathbf{w}_{\text{jun}} \\[5pt]
\mathbf{w}_{\text{in}}
\end{pmatrix}.
\]

Dividing Equation (\ref{eq:variaceTwoLevels}) by the normalization constant, the portfolio variance decomposes as:
\[
\langle \mathbf{w}^{*}, \mathbf{w}^{*} \rangle_{\Sigma} =
\langle \mathbf{w}^{*}_{\text{jun}}, \mathbf{w}^{*}_{\text{jun}} \rangle_{\mathbf{S}} 
+ \frac{\mathbf{1}_{\text{in}}^t \mathbf{X}^{-1} \mathbf{1}_{\text{in}}}{(\mathbf{1}_{\text{jun}}^t \mathbf{w}_{\text{jun}} + \mathbf{1}_{\text{in}}^t \mathbf{w}_{\text{in}})^2}.
\]
\end{enumerate}
\end{theorem}

\begin{proof}
The proof follows directly from the application of Propositions \ref{prop:Schur} and \ref{prop:DecompOfVariance}, where the vector \( \mathbf{y} \) in both cases is given by \( \mathbf{y} = \mathbf{1} \). The identity (\ref{eq:variaceTwoLevels}) follows from Proposition \ref{prop:DecompOfVariance}, where the last two terms vanish due to the condition \( \mathbf{w}_{\text{in}} = h(\mathbf{w}_{\text{jun}}, \mathbf{1}) \).
\end{proof}
One key advantage of dividing assets into junction assets and interior assets is that it significantly improves the efficiency of computing the inverse of the covariance matrix. As outlined in Theorem \ref{thm:mainResult}, Steps 1, 2, and 3, instead of inverting the full covariance matrix \( \mathbf{\Sigma} \), we only need to compute the inverses of the submatrices \( \mathbf{X} \) and \( \mathbf{S}(\Sigma) \). Notably, the inverse of \( \mathbf{X} \) is particularly straightforward due to its block diagonal structure:

\[
\mathbf{X}^{-1} =
  \setlength{\arraycolsep}{0pt}
  \setlength{\delimitershortfall}{0pt}
  \begin{pmatrix}
    \,\fbox{$B^{-1}_1$} & 0 & 0 & \cdots & 0\,  \\
    \,0 & \fbox{$B^{-1}_2$} & 0 & \cdots & 0\,  \\
    \,0 & 0 & \ddots & \ddots & \vdots\,  \\
    \,\vdots & \vdots & \ddots & \ddots & 0\,  \\
    \,0 & 0 & \cdots & 0 & \fbox{$B^{-1}_m$}\,  
  \end{pmatrix}
\]
\\
where each block corresponds to a matrix of size \( 3 \times 3 \), namely the covariance matrix of assets within a single cluster (within a copy of $G_1$), as seen in (\ref{eq:SigmaLevel2}). The number of blocks matches the number of clusters. This block-wise inversion significantly reduces computational complexity compared to directly inverting \( \mathbf{\Sigma} \). However, the situation with \( \mathbf{S}(\Sigma) \) is more intricate, as \( \mathbf{S}(\Sigma) \) is obtained via the Schur complement, capturing the interactions between junction and interior nodes. To better understand the role and behavior of \( \mathbf{S}(\Sigma) \), we now examine a concrete example.

\begin{example}[Continuation of Example \ref{ex:Level2Example1}]
\label{ex:Level2Example2}
From this point forward, we emphasize the level at which we are working, as this distinction will soon become important. Applying Theorem \ref{thm:mainResult}, we explicitly compute the optimal weight vector \( \mathbf{w}^*_2 \) (where the subscript denotes level 2 of the Sierpinski graph) and compare it with the previously derived results in (\ref{eq:optimalWeightsConcrete}). The first step is to compute the Schur complement of (\ref{eq:SigmaLevel2}):

\begin{equation}
\label{eq:SchurComp}
\renewcommand{\arraystretch}{1.5} 
\mathbf{S}(\Sigma_2) =
\begin{pmatrix}
\begin{array}{ccc|ccc}
\frac{1051}{181} & 0 & 0 & \frac{288}{181} & -\frac{57}{362} & 0\\
0 & \frac{2438}{209} & 0 & -\frac{195}{418} & 0 & \frac{117}{418}\\
0 & 0 & \frac{4289}{733} & 0 & \frac{110}{733} & \frac{60}{733}\\
\hline
\frac{288}{181} & -\frac{195}{418} & 0 & \frac{518273}{151316} & \frac{38}{181} & -\frac{107}{836}\\
-\frac{57}{362} & 0 & \frac{110}{733} & \frac{38}{181} & \frac{3010675}{265346} & -\frac{74}{733}\\
0 & \frac{117}{418} & \frac{60}{733} & -\frac{107}{836} & -\frac{74}{733} & \frac{4124565}{612788}
\end{array}
\end{pmatrix}
\renewcommand{\arraystretch}{1} 
\end{equation}
\\
By inverting this matrix, we obtain the junction weights \( \mathbf{w}_{2,\text{jun}} \) as specified in step one of the theorem:
\begin{equation}
\label{eq:w_jun2}
\mathbf{w}_{2,\text{jun}} \approx 
\left( 0.1, \  0.087, \  0.241, \  0.393, \  0.078, \  0.16 \right)^t
\end{equation}
Once the junction weights are established, the interior node weights follow from step two of the theorem:
\begin{equation}
\label{eq:w_in2}
\mathbf{w}_{2,\text{in}} \approx 
\left( -0.101, \  0.195, \  0.208, \  0.26, \  0.283, \  0.218, \  0.074, \  0.186, \  0.1 \right)^t
\end{equation}
With these components in place, we ensure normalization by computing:
\begin{equation*}
\mathbf{1}_{\text{jun}}^t \mathbf{w}_{2,\text{jun}}  + \mathbf{1}_{\text{in}}^t \mathbf{w}_{2,\text{in}} \approx  2.481
\end{equation*}
This leads to the final optimal weight vector \( \mathbf{w}^*_2 \), which matches the previously derived result in (\ref{eq:optimalWeightsConcrete}). The next step is to decompose the portfolio variance using Theorem \ref{thm:mainResult}. Specifically, the variance contribution from the junction weights at level 2 is:
\begin{equation}
\langle \mathbf{w}^{*}_{2,\text{jun}}, \mathbf{w}^{*}_{2,\text{jun}} \rangle_{\mathbf{S}(\Sigma_2)} \approx  0.221
\end{equation} 
The total portfolio variance is then computed as
$\langle \mathbf{w}^{*}_2, \mathbf{w}^{*}_2 \rangle_{\Sigma} \approx 0.403$, which also coincides with the previous result in (\ref{eq:optimalWeightsConcrete}).
\end{example}
\begin{figure}[!htb]
        \centering     
   		\includegraphics[scale=0.85]{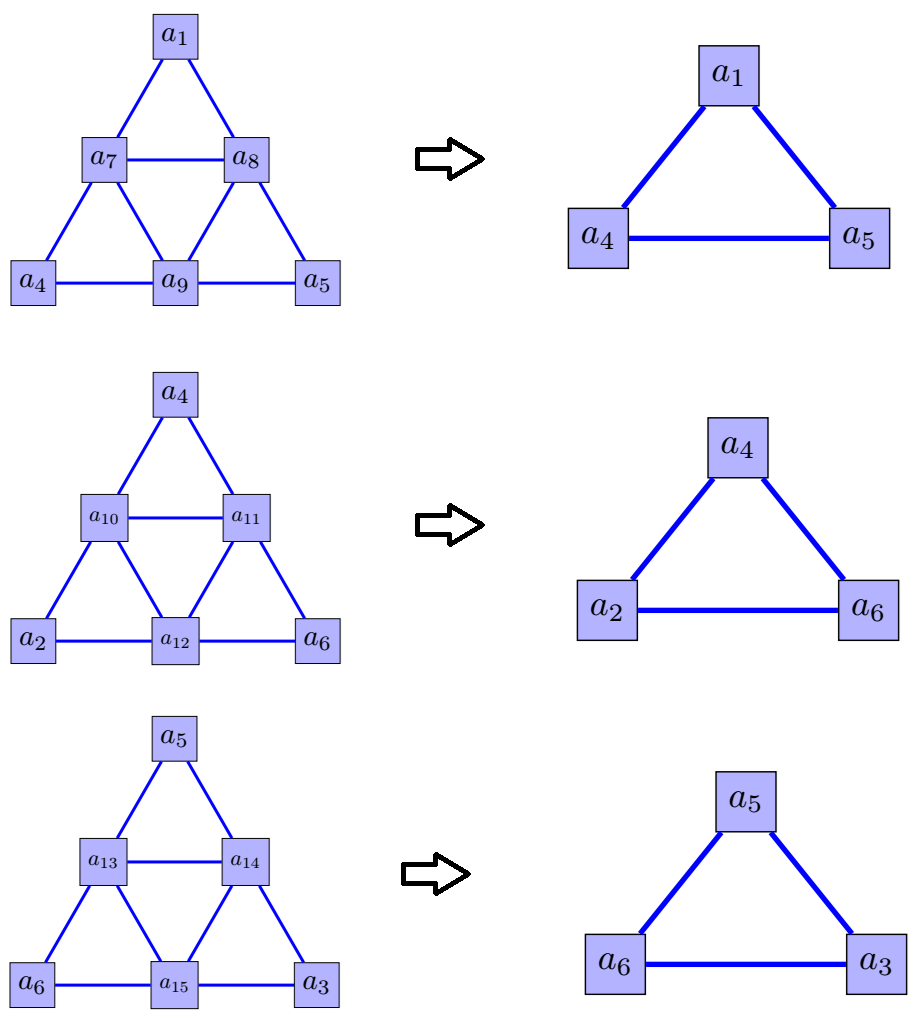}
    	\caption{Intuitive illustration of the Schur complement. The Schur complement acts as a transformation that reduces the complexity of the covariance structure by integrating interior assets within each cluster. This effectively replaces them with an adjusted covariance among the corresponding junction assets. For example, in the top cluster, assets \(a_1\), \(a_4\), and \(a_5\) are initially uncorrelated. However, after applying the Schur complement, these assets become correlated. This correlation is artificial, accounting for the influence of the interior assets.}
	\label{fig:trafoFromL2toL1}
\end{figure}
Equation (\ref{eq:SchurComp}) reveals an essential interpretation: 
$\mathbf{S}(\Sigma_2)$ represents the covariance matrix of a level-1 portfolio. This portfolio follows the structure of $G_1$, the leftmost graph in Figure \ref{fig:sierpinski_graphs}, 
and the corresponding generic covariance matrix has the following form,

\begin{align}
\label{eq:Sigma1Generic}
\mathbf{\Sigma}_1 
=
\begin{pmatrix}
T_1 & J_1^t \\[5pt]
J_1 & X_1
\end{pmatrix}
=
\begin{pmatrix}
\begin{array}{ccc|ccc}
\sigma_{11} & 0 & 0 & \sigma_{14} & \sigma_{15} & 0 \\ 
0 & \sigma_{22} & 0 & \sigma_{24} & 0 & \sigma_{26} \\ 
0 & 0 & \sigma_{33} & 0 & \sigma_{35} & \sigma_{36} \\ 
\hline
\sigma_{14} & \sigma_{24} & 0 & \sigma_{44} & \sigma_{45} & \sigma_{46} \\ 
\sigma_{15} & 0 & \sigma_{35} & \sigma_{45} & \sigma_{55} & \sigma_{56} \\ 
0 & \sigma_{26} & \sigma_{36} & \sigma_{46} & \sigma_{56} & \sigma_{66}
\end{array}
\end{pmatrix}
\end{align}
\\
To understand this observation, let's provide an intuitive explanation for the Schur complement. The level-2 portfolio, shown in Figure \ref{fig:portfolioLevel2}, consists of three clusters (copies of \( G_1 \)), which have been separated in Figure \ref{fig:trafoFromL2toL1} for clarity. When we apply the Schur complement at level 2, each cluster is transformed into a triangular graph, forming an equivalent portfolio that consists only of the junction assets. For instance, in the top cluster of Figure \ref{fig:trafoFromL2toL1}, assets \( a_1 \), \( a_4 \), and \( a_5 \) are initially uncorrelated, as they are not directly connected by edges in level 2. However, after applying the Schur complement, these assets become correlated, as they are now connected by edges in the resulting triangular graph. This correlation is not intrinsic but emerges as a result of integrating out the interior assets. To account for the influence of these interior assets, the junction assets are treated as correlated in level 1, even though they were not directly correlated in level 2. This effect is a fundamental property of the Schur complement and holds at all levels of the hierarchy.

\begin{figure}[!htb]
        \centering     
        \includegraphics[scale=0.7]{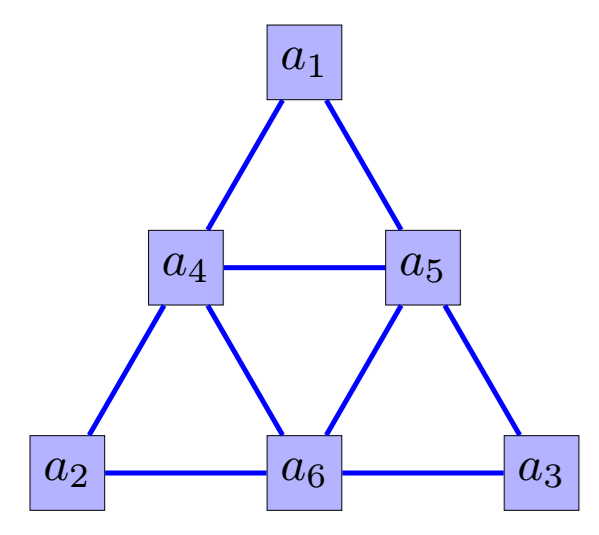}
    	\caption{After applying the Schur complement to each of the three clusters in Figure \ref{fig:trafoFromL2toL1}, the interior assets are eliminated, and the remaining junction assets form three triangular graphs. These three triangles are then merged to obtain the level-1 graph \( G_1 \), effectively reducing the hierarchical level.}

	\label{fig:portfolioLevel1}
\end{figure}

When the three triangular graphs obtained through the Schur complement are merged, they form the level-1 graph, representing the new effective portfolio (see Figure \ref{fig:portfolioLevel1}). This transformed portfolio consists only of the junction assets of level 2, while the interior assets have been integrated out. In this effective portfolio, \( a_1, a_2, \) and \( a_3 \) become the new junction assets, while \( a_4, a_5, \) and \( a_6 \) serve as the interior assets. The covariance matrix of the new portfolio is given by \( \Sigma_1 = \mathbf{S}(\Sigma_2)  \) and has the form (\ref{eq:Sigma1Generic}).

While this explanation uses a specific example, the same process applies at any hierarchical level. More generally, if we start at level \( \ell \) and systematically integrate the interior assets within each copy of \( G_1 \), we obtain the graph of level \( \ell -1 \), which consists only of the junction assets of level $\ell$. Effectively, the Schur complement reduces the hierarchical level by one. This operation systematically eliminates the interior assets while preserving their influence through an adjusted covariance structure among the remaining junction assets. In this sense, applying the Schur complement corresponds to incrementing the hierarchical level by \(-1\).

Now, given the role of the Schur complement in a hierarchical structure, we revisit Example \ref{ex:Level2Example2} and take a different approach. Instead of directly solving for \( \mathbf{w}_2 \) using the inverse of \( \mathbf{S}(\Sigma_2) \), we adopt an iterative method, applying the Schur complement at each hierarchical level. In other words, we iteratively apply Theorem \ref{thm:mainResult} and Proposition \ref{prop:Schur} across different hierarchical levels. This requires modifying the notation in these results to accommodate computations across multiple levels.

\begin{definition}
\label{def:MainDef}
Suppose we construct a portfolio at level \( \ell \), where the covariance matrix is:
\[
\Sigma_{\ell} = \text{Portfolio Covariance Matrix}.
\]
To systematically reduce complexity, we reorder the assets by grouping junction nodes before interior nodes, processing them cluster by cluster. This structured ordering allows us to decompose \( \Sigma_{\ell} \) into block matrices and iteratively apply the Schur complement, obtaining an effective covariance matrix at the next lower level, \( \Sigma_{\ell-1} \). Repeating this process recursively yields the covariance matrix at level \( k \) as follows:
\begin{enumerate}
\item  The covariance matrix at level \( k \):
\begin{equation*}
\Sigma_k = S(\Sigma_{k+1}) = T_{k+1} - \mathbf{J}_{k+1}^t \mathbf{X}_{k+1}^{-1} \mathbf{J}_{k+1}, \quad \quad k = 0, \dots, \ell-1.
\end{equation*}
    \item For \( k = 1, \dots, \ell \), let \( \mathbf{y}_k \) be a vector defined on all nodes at level \( k \), and let \( \mathbf{v}_k \) be a vector defined on the junction nodes of level \( k \). The function \( h_k \) is given by:
    \begin{equation*}
        h_k(\mathbf{v}_k, \mathbf{y}_k) = \mathbf{X}_k^{-1} \mathbf{y}_{k,\text{in}} - \mathbf{X}_k^{-1} \mathbf{J}_k \mathbf{v}_k.
    \end{equation*}
    This function computes a vector defined on the interior nodes of level \( k \).

    \item For \( k = 1, \dots, \ell \), the function \( g_{k-1} \) maps a vector \( \mathbf{y}_k \), defined on all nodes at level \( k \), to a new vector at level \( k-1 \):
    \begin{equation*}
        g_{k-1}(\mathbf{y}_k) = \mathbf{y}_{k,\text{jun}} - \mathbf{J}_k^t \mathbf{X}_k^{-1} \mathbf{y}_{k,\text{in}}.
    \end{equation*}

    \item To simplify notation, we define the following recursive sequence for \( k = 0, \dots, \ell \):
    \begin{equation*}
        \gamma_{\ell} = \mathbf{1}, \quad 
        \gamma_{\ell-1} = g_{\ell-1}(\gamma_{\ell}), \quad \dots, \quad 
        \gamma_0 = g_0(\gamma_1).
    \end{equation*}
\end{enumerate}

\end{definition}

\begin{theorem}
\label{thm:mostimportant}
Let \( 0 \leq k-1 \leq k \leq \ell \) and  $\gamma_k, h_k, g_k$ are given as in Definition \ref{def:MainDef}. A vector \( \mathbf{w}_k \) satisfies the equation $\mathbf{w}_k = \Sigma_k^{-1} \gamma_k$
if and only if the vector \( \mathbf{w}_{k-1} \) satisfies:
$\mathbf{w}_{k-1} = \Sigma_{k-1}^{-1} \gamma_{k-1}$, where $\gamma_{k-1} = g_{k-1}(\gamma_k)$ and \( \mathbf{w}_k \) can be explicitly written in terms of \( \mathbf{w}_{k-1} \) as:
\[
\mathbf{w}_k =
\begin{pmatrix}
\mathbf{w}_{k-1} \\[5pt]
h_k(\mathbf{w}_{k-1}, \gamma_k)
\end{pmatrix}
\]
\end{theorem}

\begin{example}[Continuation of Example \ref{ex:Level2Example2}]
We revisit Example \ref{ex:Level2Example2}, now applying the Schur complement iteratively at each level using Theorem \ref{thm:mostimportant}, following the notation introduced in Definition \ref{def:MainDef}.

\begin{enumerate}
    \item \textbf{Initialize at Level 2}:
    \begin{itemize}
        \item We initialize: $\ \gamma_2 = \mathbf{1}$, and  $\mathbf{w}_2 = \Sigma_2^{-1} \gamma_2$
    \end{itemize}

    \item \textbf{Decompose weights at Level 2}:
    \begin{itemize}
        \item We decompose \( \mathbf{w}_2 \) as:
        \[
        \mathbf{w}_2 =
        \begin{pmatrix}
        \mathbf{w}_1 \\[5pt]
        h_2(\mathbf{w}_1, \gamma_2)
        \end{pmatrix}
        \]
        \item Since \( \mathbf{w}_1 = \mathbf{w}_{2,\text{jun}} \), we apply $$ \mathbf{w}_{2,\text{in}}=h_2(\mathbf{w}_1, \gamma_2) = \mathbf{X}_2^{-1} \gamma_{2, \text{in}} - \mathbf{X}_2^{-1} \mathbf{J}_2 \mathbf{w}_1$$
    \end{itemize}

    \item \textbf{Reduce to Level 1}:
    \begin{itemize}
        \item The weight vector at level 1 satisfies:
        \[
        \mathbf{w}_1 = \Sigma_1^{-1} \gamma_1, \quad \quad \gamma_1 = g_1(\gamma_2)
        \]
    \end{itemize}

 \item \textbf{Compute \( \gamma_1 \)}:
\begin{itemize}
\item  Using Theorem \ref{thm:mostimportant}, we obtain: $$ \gamma_1 = g_1(\gamma_2) = \gamma_{2, \text{jun}} - \mathbf{J}_2^t \mathbf{X}_2^{-1} \gamma_{2, \text{in}}$$
    \item The vector \( \gamma_2 = \mathbf{1} \) at level 2 consists of 15 entries, all equal to 1. Since there are 6 junction nodes (matching the number of nodes in level 1), the first 6 entries form \( \mathbf{1}_{\text{jun}} \), while the remaining 9 entries form \( \mathbf{1}_{\text{in}} \).
    
    \item Using the block matrices \( \mathbf{J}_2^t \) and \( \mathbf{X}_2 \) from (\ref{eq:SigmaLevel2}), we compute:
    \[
    \gamma_1 \approx \begin{pmatrix} 1.191,\ 0.871,\ 1.437,\ 1.459,\ 0.976,\ 1.06 \end{pmatrix}^t
    \]
\end{itemize}
   
    \item \textbf{Decompose weights at Level 1}:
    \begin{itemize}
        \item We decompose:
        \[
        \mathbf{w}_1 =
        \begin{pmatrix}
        \mathbf{w}_0 \\[5pt]
        h_1(\mathbf{w}_0, \gamma_1)
        \end{pmatrix}.
        \]
        \item Since \( \mathbf{w}_0 = \mathbf{w}_{\text{jun},1} \), we apply:
        \[
        \mathbf{w}_{\text{in},1} = h_1(\mathbf{w}_0, \gamma_1) = \mathbf{X}_1^{-1} \gamma_{1, \text{in}} - \mathbf{X}_1^{-1} \mathbf{J}_1 \mathbf{w}_0.
        \]
         \item The term $\gamma_{1, \text{in}}$ represents the interior components of the vector \( \gamma_{1} \), which is decomposed as:
    \begin{equation*}
    \gamma_{1} = (\gamma_{1, \text{jun}},\gamma_{1, \text{in}})^t
    \end{equation*}
 
    \item Note that \( \gamma_{1} \) is a vector on the nodes of \( G_1 \). Since \( G_1 \) has three junction and three interior nodes, we extract its junction and interior components from Step 4:
    \[
    \gamma_{1, \text{jun}} \approx 
    \begin{pmatrix} 1.191, 0.871, 1.437 \end{pmatrix}^t, \quad \quad
    \gamma_{1, \text{in}} \approx 
    \begin{pmatrix} 1.459, 0.976, 1.06 \end{pmatrix}^t
    \]
    \end{itemize}

    \item \textbf{Reduce to Level 0}:
    \begin{itemize}
        \item The weight vector at level 0 satisfies:
        \[
        \mathbf{w}_0 = \Sigma_0^{-1} \gamma_0, \quad \quad \gamma_0 = g_0(\gamma_1).
        \]
    \end{itemize}

    \item \textbf{Compute \( \gamma_0 \)}:
    \begin{itemize}
        \item Using Theorem \ref{thm:mostimportant}, we obtain:
        \[
        \gamma_0 = g_0(\gamma_1) = \gamma_{1, \text{jun}} - \mathbf{J}_1^t \mathbf{X}_1^{-1} \gamma_{1, \text{in}}.
        \]
    \item Substituting the block matrices \( \mathbf{J}_1^t \) and \( \mathbf{X}_1 \) from (\ref{eq:SchurComp}) and the junction and interior values of \( \gamma_1 \) from Step 5, we compute:
    \[
    \gamma_0 \approx \left( 0.523, \  1.023, \  1.411\right)^t
    \]
    \end{itemize}

    \item \textbf{Final Covariance Matrix at Level 0}:
\begin{itemize}
    \item By computing the Schur complement using the block matrices from (\ref{eq:SchurComp}), we derive the effective covariance matrix at the base level:

    \[
    \Sigma_{0} \approx
    \begin{pmatrix}
    5.061 & 0.215 & 0.003\\
    0.215 & 11.591 & -0.004\\
    0.003 & -0.004 & 5.848
    \end{pmatrix}.
    \]
    \\
\end{itemize}

    \item \textbf{Compute Portfolio Weights at Each Level}:
    \begin{itemize}
        \item We recursively compute:
        
        \[
                \mathbf{w}_2 =
        \begin{pmatrix}
        \mathbf{w}_1 \\[5pt]
        h_2(\mathbf{w}_1, \gamma_2)
        \end{pmatrix}
         \quad \rightarrow \quad
        \mathbf{w}_1 =
        \begin{pmatrix}
        \mathbf{w}_0 \\[5pt]
        h_1(\mathbf{w}_0, \gamma_1)
        \end{pmatrix}
        \quad \rightarrow \quad
\mathbf{w}_0 = \Sigma_0^{-1} \gamma_0
        \]
\\
    \item Using \( \Sigma_0 \) from Step 8 and \( \gamma_0 \) from Step 7, we compute \( \mathbf{w}_0 \) as described in Step 6:
    \[
    \mathbf{w}_0 \approx \left( 0.1, \  0.087, \  0.241\right)^t
    \]
    These values correspond to the weights assigned to assets \( a_1, a_2, \) and \( a_3 \), and they match the first three values in (\ref{eq:w_jun2}).

    \item To compute \( \mathbf{w}_1 = (\mathbf{w}_{\text{jun},1}, \mathbf{w}_{\text{in},1})^t \), we follow the procedure from Step 5. Recall that \( \mathbf{w}_{\text{jun},1} \) is equal to \( \mathbf{w}_0 \), which we have just computed. Using the formula from Step 5, we obtain:
    
\[
\mathbf{w}_{\text{in},1} \approx  \left( 0.393, \  0.078, \  0.16\right)^t
\]
These values correspond to the weights assigned to assets \( a_4, a_5, \) and \( a_6 \), and they match the fourth, fifth, and sixth entries in (\ref{eq:w_jun2}). 

    \item To compute \( \mathbf{w}_2 = (\mathbf{w}_{\text{jun},2}, \mathbf{w}_{\text{in},2})^t \), we follow the procedure from Step 2. Recall that \( \mathbf{w}_{\text{jun},2} \) is equal to \( \mathbf{w}_1 \), which we have just computed. Using the formula from Step 2, we obtain:
\[
\mathbf{w}_{\text{in},2} \approx \left( -0.101, \  0.195, \  0.208, \  0.26, \  0.283, \  0.218, \  0.074, \  0.186, \  0.1\right)^t
\]
These values correspond to the weights assigned to assets \( a_7 \) through \( a_{15} \) and match the respective entries in (\ref{eq:w_in2}).
\end{itemize}
\end{enumerate}
\end{example}
In the steps above, we only inverted \( 3 \times 3 \) matrices, specifically the block matrices in \( \mathbf{X}_k \) and \( \Sigma_0 \), never larger ones. These computations form the first two steps of an algorithm that, in the next section, is generalized to initialize a portfolio at level \( n \).

\section{Hierarchical Minimum Variance Portfolios (HMVP) Algorithm}
\label{sec:Algo}

This section presents a Hierarchical Minimum Variance Portfolios (HMVP) Algorithm, a recursive method for constructing minimum-variance portfolios using hierarchical clustering and covariance decomposition.

\subsection{Algorithm Steps}

\subsubsection{Step 1: Initialization}
A portfolio of $n$ assets is considered, where the covariance matrix $\Sigma(\ell)$ can be structured as a hierarchical graph of level $\ell$.

\textbf{Input:}
\begin{itemize}
\item  $\{ {G}_i \}_{i \in \nn}$: A class of hierarchical graphs used to structure the portfolio. 
\item $\ell$: Maximum hierarchical level.
\item $\Sigma(\ell)$: Covariance matrix of $n$ assets, structured as a hierarchical graph of level $\ell$.
    
\end{itemize}

\textbf{Output:}
\begin{itemize}
    \item Optimal portfolio weight vector $w^*$.
\end{itemize}

\begin{remark}
To construct \( \Sigma_\ell \), we first select a hierarchical graph class. Suppose we choose the family of Sierpinski graphs. The next step is to determine the appropriate level \( \ell \), which depends on the number of assets we want to include in the portfolio. A Sierpinski graph of level \( \ell \) consists of  

\[
N_\ell = \frac{1}{2} (3^{\ell+1} + 3)
\]
nodes, while the number of junction nodes at this level matches the total number of nodes in a Sierpinski graph of level \( \ell-1 \), given by  

\[
J_\ell = \frac{1}{2} (3^\ell + 3).
\]
For example, at level \( \ell = 1 \), the graph has 6 nodes, including 3 junction nodes. At level \( \ell = 2 \), the number of nodes increases to 15, with 6 junction nodes. Similarly, at \( \ell = 3 \), the graph contains 42 nodes and 15 junction nodes, while at \( \ell = 4 \), it expands to 123 nodes with 42 junction nodes. 

Now, suppose we decide on \( \ell = 4 \). In this case, we construct a basket of 42 uncorrelated assets, which serve as the junction nodes and form the foundation for diversification. We then assign these 42 assets to the junction nodes of \( G_4 \). After that, we begin constructing clusters by identifying groups of three correlated assets that connect to three of the junction assets. This process ensures that the hierarchical structure captures the underlying correlation patterns in the portfolio. To scale this approach efficiently, we need to automate the procedure, ensuring that asset selection and cluster formation lead to optimized outcomes.

\end{remark}

\subsubsection{Step 2:}
At each level $k$, the covariance matrix $\Sigma(k)$ is decomposed into junction and interior components, and the Schur complement is computed to derive a reduced covariance matrix for the next level.

\begin{algorithm}[H]
\caption{Recursive Covariance Matrix Reduction}
\begin{algorithmic}
\For{each level $k$ from $\ell$ down to 1}
    \State Decompose covariance matrix $\Sigma(k)$ into:
        \begin{itemize}
            \item $T(k) \leftarrow$ Covariance submatrix of junction assets.
            \item $X(k) \leftarrow$ Covariance submatrix of interior assets.
            \item $J(k) \leftarrow$ Interaction matrix between junction and interior assets.
        \end{itemize}
    \State Compute Schur complement:
        \[
        \Sigma(k-1) \leftarrow T(k) - J(k)^T X(k)^{-1} J(k)
        \]
\EndFor
\end{algorithmic}
\end{algorithm}

\subsubsection{Step 3: Structural Mappings}

At each hierarchical level, assets are categorized into junction and interior groups.

\begin{algorithm}[H]
\caption{Partition Function}
\begin{algorithmic}[1]
\Function{$partition$}{$k, y$} \\
    \textbf{Input:} 
    \begin{itemize}
        \item $k$: hierarchical level.
        \item $y$: vector defined on all nodes of level $k$.
    \end{itemize}
    \textbf{Output:} 
    \begin{itemize}
        \item $y_{\text{jun}}(k)$: values of $y$ corresponding to junction nodes.
        \item $y_{\text{in}}(k)$: values of $y$ corresponding to interior nodes.
    \end{itemize}
    \text{ }
    \State
    Extract $y_{\text{jun}}(k)$ and $y_{\text{in}}(k)$ from $y$ 
    \State \Return $\ y_{\text{jun}}(k), \ y_{\text{in}}(k)$
\EndFunction
\end{algorithmic}
\end{algorithm}

\noindent The following pseudocode computes \( g_k(y_{k+1}) \) from Definition \ref{def:MainDef}.

\begin{algorithm}[H]
\caption{Computation of $g(k, y)$}
\begin{algorithmic}[1]
\Function{$g$}{k, y}
    \\ \textbf{Input:} 
    \begin{itemize}
        \item $k$: hierarchical level.
        \item $y$: vector defined on all nodes of level $k+1$.
    \end{itemize}
    \textbf{Output:} 
    \begin{itemize}
        \item A vector defined on junction nodes at level $k$.
    \end{itemize}
    \text{ }
    \State $y_{\text{jun}}(k+1)$, $ \ y_{\text{in}}(k+1)$ $\leftarrow$  $partition$($k+1, y$)
    \State Compute:
        \[
        g(k, y) \leftarrow y_{\text{jun}}(k+1) - J(k+1)^T X(k+1)^{-1} y_{\text{in}}(k+1)
        \]
    \State \Return $g(k, y)$
\EndFunction
\end{algorithmic}
\end{algorithm}

\noindent The vector $\gamma(k)$ plays the role of vector $\mathbf{1}$ from the original minimum-variance optimization problem, transforming it through the hierarchy to adjust portfolio weights consistently at each level.

\begin{algorithm}[H]
\caption{Recursive Computation of $\gamma(k)$}
\begin{algorithmic}[1]
\Function{$\gamma$}{$k$} \\
    \textbf{Input:}
    \begin{itemize}
        \item $k$: hierarchical level.
    \end{itemize}
    \textbf{Output:}
    \begin{itemize}
        \item $\gamma(k)$: recursively computed vector representing transformed $\mathbf{1}$ at level $k$.
    \end{itemize}
    \If{$k == \ell$}
        \State $\gamma(k) \leftarrow 1$ (vector of ones at level $\ell$)
    \Else
        \State $\gamma(k) \leftarrow g(k, \gamma(k+1))$
    \EndIf
    \State \Return $\gamma(k)$
\EndFunction
\end{algorithmic}
\end{algorithm}

\noindent The following function recursively computes the portfolio weights from level $k=0$ to $k=\ell$. 
At the base level, it initializes the weights using the covariance matrix $\Sigma(0)$ and 
the vector $\gamma(0)$. For higher levels, it recursively computes weights from previous levels 
and updates them using the function $h(k, \cdot, \cdot)$.

\begin{algorithm}[H]
\caption{Recursive Computation of Portfolio Weights}
\begin{algorithmic}[1]
\Function{$compute$\_$weights$}{$k$} \\
    \textbf{Input:} 
    \begin{itemize}
        \item $k$: Hierarchical level.
    \end{itemize}
    \textbf{Output:} 
    \begin{itemize}
        \item $w(k)$: Not normalized portfolio weight vector at level $k$.
    \end{itemize}
    \text{ }
    \If{$k == 1$} 
        \State Compute:
        \[
        \begin{aligned}
        w_{\text{jun}}(1) \ &\leftarrow \ \Sigma(0)^{-1} \gamma(0) \\
        w_{\text{in}}(1) \ &\leftarrow \ h(1, w_{\text{jun}}(1), \gamma(1)) \\
        w(1)          \    &\leftarrow \ (w_{\text{jun}}(1), w_{\text{in}}(1))
        \end{aligned}
        \]
        \State \Return $w(1)$
    \Else
        \State Compute:
        \[
        \begin{aligned}
        w(k-1)           &\leftarrow \text{$compute$\_$weights$}(k-1) \\
        w_{\text{jun}}(k) &\leftarrow w(k-1) \\
        w_{\text{in}}(k)  &\leftarrow h(k, w_{\text{jun}}(k), \gamma(k)) \\
        w(k)              &\leftarrow (w_{\text{jun}}(k), w_{\text{in}}(k))
        \end{aligned}
        \]
        \State \Return $w(k)$
    \EndIf
\EndFunction
\end{algorithmic}
\end{algorithm}

\begin{algorithm}
\caption{Normalization of Portfolio Weights}
\begin{algorithmic}[1]
\Function{normalize\_weights}{$w$}
    \State Compute:
        \[
        w^* \leftarrow \frac{w}{\sum w}
        \]
    \State \Return $w^*$
\EndFunction
\end{algorithmic}
\end{algorithm}

\section{Conclusion and Future Research}

In this work, we introduce a novel Hierarchical Minimum Variance Portfolio (HMVP) approach, which leverages hierarchical graph structures to optimize portfolio allocation within the framework of Markowitz’s mean-variance optimization, using Schur complement methods. Our approach builds upon López de Prado’s ideas and Cotton's recent work on incorporating hierarchy into portfolio construction to better reflect the underlying asset relationships and dependencies.

While López de Prado’s Hierarchical Risk Parity (HRP) method organizes assets using a tree-based structure, our approach differs by employing hierarchical graphs. One advantage of this method is that it allows for a structured hierarchy of graphs, where complexity increases systematically in the sequence of graphs as each level is constructed by combining multiple copies of the previous level. This property is directly reflected in the portfolio construction process: if a portfolio’s covariance structure follows a graph within a hierarchical sequence, we can reduce complexity by iteratively simplifying the graph structure down to its fundamental building block. Importantly, this hierarchical reduction retains full covariance information even as complexity is reduced. A similar concept was noted in Cotton’s work in a different setting, where assets were divided into two groups, inducing a block structure and applying the Schur complement. In our approach, this reduction emerges naturally from the graph structure itself. By analyzing node properties, we distinguish two types: junction nodes and interior nodes, which correspond to different asset roles within the hierarchical structure. This distinction leads to a particular block matrix decomposition and a Schur complement formulation. In our framework, the Schur complement serves as a fundamental mechanism that enables transitions between successive levels in the hierarchy while preserving structural consistency and covariance relationships at each level. One of our main results, Theorem \ref{thm:mostimportant}, establishes an explicit relationship between the portfolio weight vectors \(w_k\) and \(w_{k-1}\) at two successive levels in the hierarchy. This result provides the foundation for constructing our algorithm recursively.

Moreover, unlike HRP, our method does not require transforming correlation data into a distance metric. Instead, we model portfolios using weighted hierarchical graphs, where edges represent raw covariance data. In Theorem \ref{thm:mainResult}, we establish Cotton's idea of unifying hierarchical portfolio methods, the Markowitz minimum variance portfolio, and the Schur complement by demonstrating that portfolio optimization can be systematically reduced. Specifically, we show that the reduced optimization problem corresponds to an effective covariance matrix obtained through the Schur complement. Furthermore, our approach does not require the inter- and intra-group allocation normalizations present in Cotton’s method, as the reduced optimization problem naturally incorporates these adjustments.

\begin{figure}[h]
    \centering
    \includegraphics[width=0.6\textwidth]{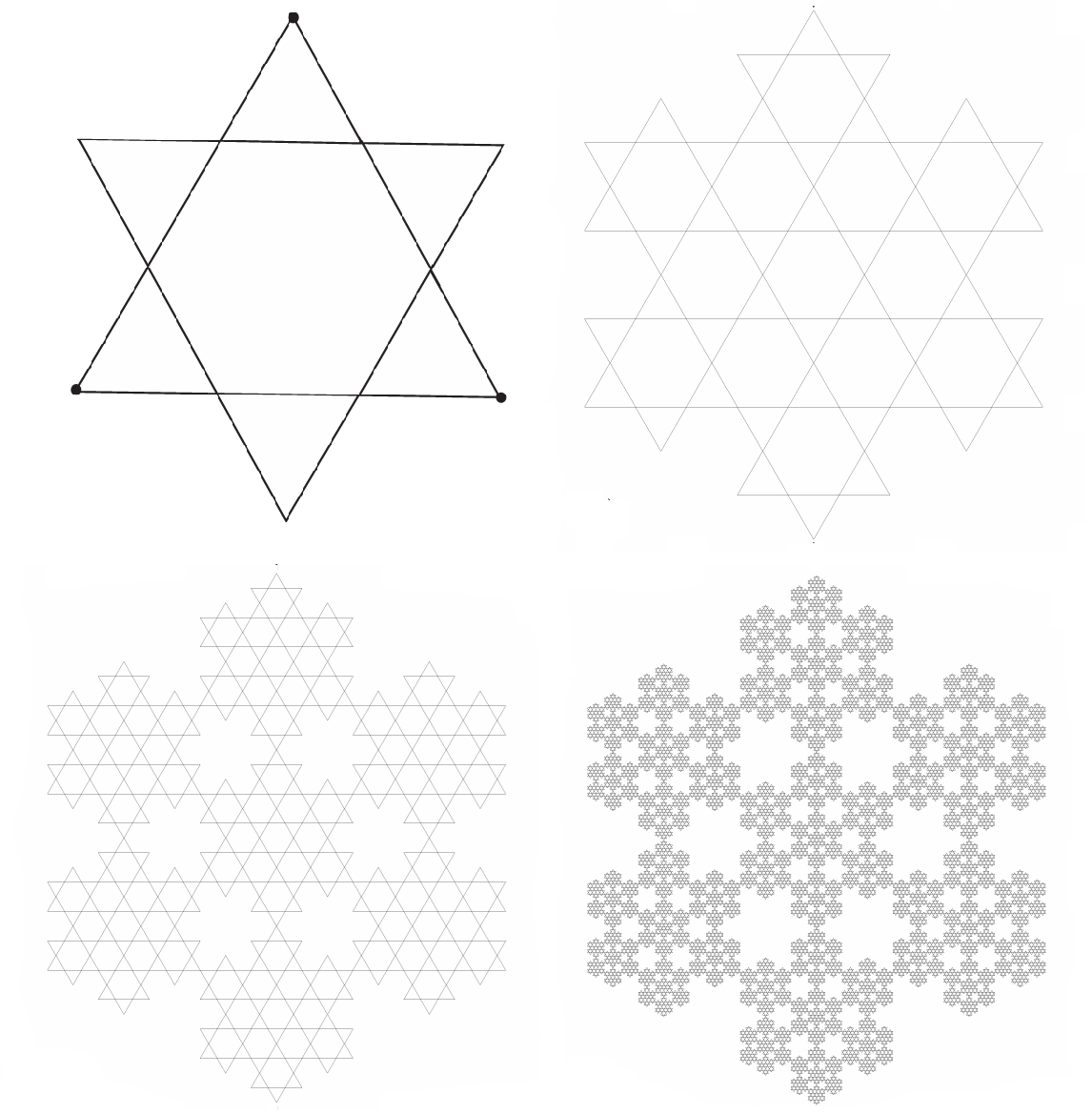}
    \caption{Several graphs from the hierarchical sequence of hexagasket-type graphs. The top-left is the base cluster, consisting of 6 junction assets and 6 internal assets. Note that the internal assets are not all correlated, although edges can be added to introduce more correlations. The top-left graph is from \cite{bajorin2008vibration}, while the other three are from \cite{tzanov2015selfsimilar}.}
    \label{fig:hexagasket}
\end{figure}

\begin{figure}[h]
    \centering
    \includegraphics[width=0.7\textwidth]{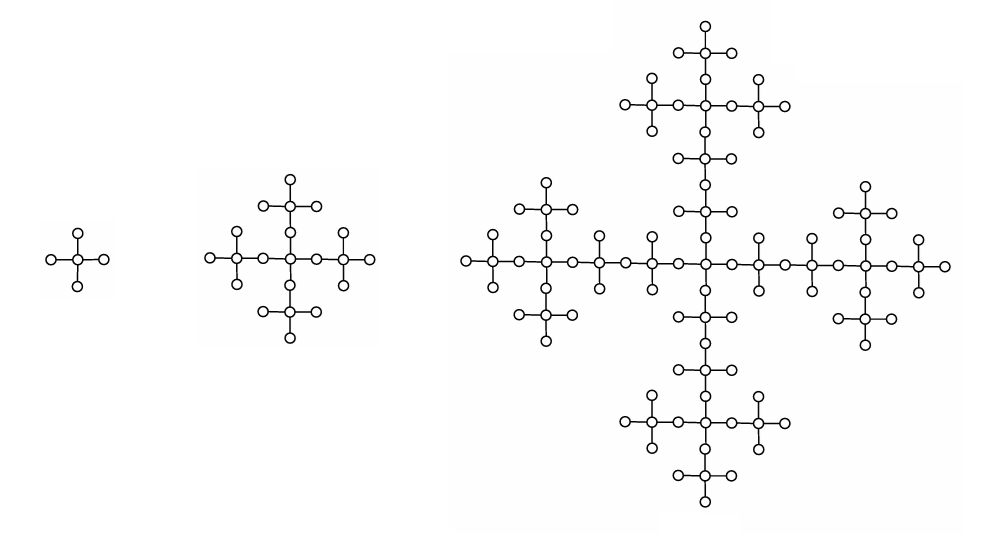}
    \caption{First three graphs of the Vicsek hierarchical graph sequence. The leftmost is the base cluster, consisting of four junction assets and one internal asset. Figure taken from \cite{patterson2011network}.}
    \label{fig:vicsek}
\end{figure}

\begin{figure}[h]
    \centering
    \includegraphics[width=0.7\textwidth]{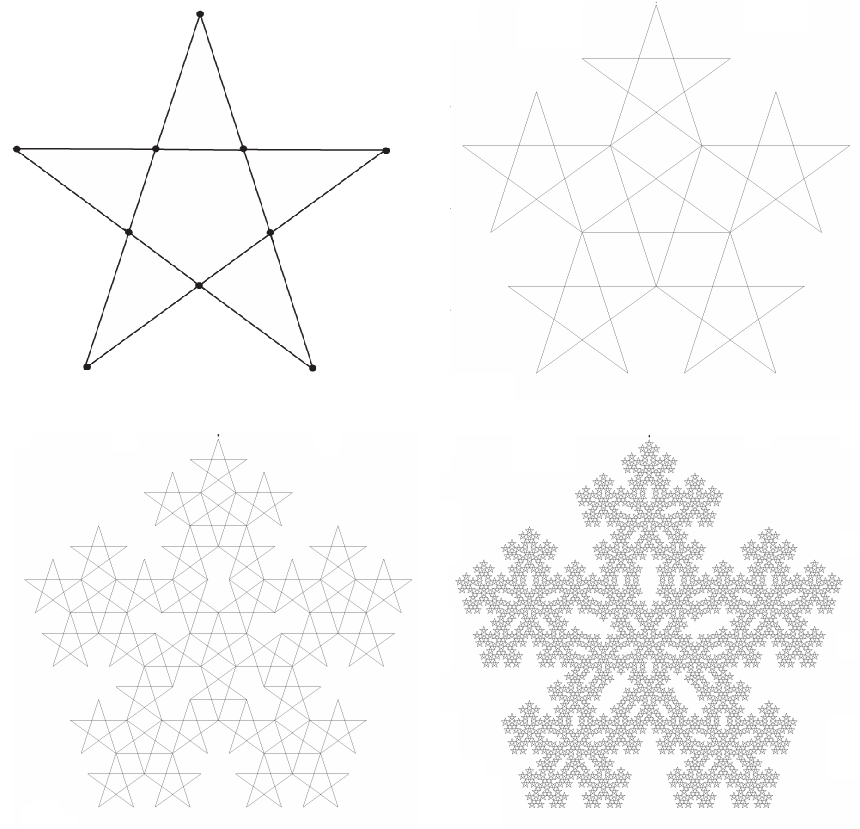}
    \caption{Several graphs from the hierarchical sequence of polygasket graphs. The top-left is the base cluster, consisting of 5 junction assets and 5 internal assets. Note that the internal assets are not all correlated, although edges can be added to introduce more correlations. The top-left graph is from \cite{strichartz2006differential}, while the other three are from \cite{tzanov2015selfsimilar}.}
    \label{fig:polygaskets}
\end{figure}

That being said, our approach has its own challenges. Not all asset relationships naturally fit into a predefined hierarchical graph, requiring careful selection or transformation of assets. While some effort may be needed to select assets that align with a hierarchical structure, the existence of numerous hierarchical graph families makes this task more feasible. A wide variety of hierarchical graphs exist, as illustrated in Figures \ref{fig:fractals}, \ref{fig:sierpinski_graphs},  \ref{fig:hexagasket}, \ref{fig:vicsek} and \ref{fig:polygaskets}. Furthermore, similar hierarchical structures could be extended into higher dimensions, incorporating greater complexity to align with the preferences of different investors or constructing hierarchies tailored to portfolio construction. A notable computational advantage of our approach is that it requires inverting only \(3 \times 3\) matrices, independent of portfolio size, to determine the optimal weights. However, like all methods reliant on covariance estimation, the accuracy of the input data remains a crucial factor.

Future research could explore ways to enhance the robustness of our approach by incorporating Bayesian estimation or machine learning techniques into graph-based methods. A promising direction is dynamic graph adaptation techniques, where the hierarchical structure evolves in response to changing market conditions, enhancing the model’s adaptability. Last but not least, we are also interested in developing a software that works with numerous families of hierarchical graphs and classifies asset baskets based on their covariance structure, fitting them into one of these hierarchical graph families.

\newpage

\end{document}